\documentstyle[12pt]{article}
\newcommand{\EPJ}{Eur. Phys. J. }

\newcommand{\JHEP}{J. High Energy Phys. }

\newcommand{\NP}{Nucl. Phys. }

\newcommand{\PRL}{Phys. Rev. Lett. }
\newcommand{\PL}{Phys. Lett. }

\begin{document}
\baselineskip=18pt

\pagenumbering{arabic}

\vspace{1.0cm}
\begin{flushright}
LU-ITP 2004/021
\end{flushright}

\begin{center}
{\Large\sf On evaluation of nonplanar diagrams in noncommutative
field theory}
\\[10pt]
\vspace{.5 cm}

{Yi Liao}
\vspace{1.0ex}

{\small Institut f\"ur Theoretische Physik, Universit\"at Leipzig,\\
Augustusplatz 10/11, D-04109 Leipzig, Germany\\}

\vspace{2.0ex}

{\bf Abstract}

\end{center}

This is a technical work about how to evaluate loop integrals appearing
in one loop nonplanar (NP) diagrams in noncommutative (NC) field theory.
The conventional wisdom says that, barring the ultraviolet/infrared
(UV/IR) mixing problem, NP diagrams whose planar counterparts are UV
divergent are rendered finite by NC phases that couple the loop momentum
to the external ones $p$ through an NC momentum
$\rho^{\mu}=\theta^{\mu\nu}p_{\nu}$.
We show that this is generally not the case. We find that
subtleties arise already in the simpler case of Euclidean spacetime.
The situation is even worse in Minkowski spacetime due to its
indefinite metric. We compare different prescriptions that may be
used to evaluate loop integrals in ordinary theory. They are
equivalent in the sense that they always yield identical results. 
However, in NC theory
there is no a priori reason that these prescriptions, except for the
defining one that is built in the Feynman propagator, are physically
justified even when they seem mathematically meaningful. Employing them
can lead to ambiguous results, which are also different from those
obtained according to the defining prescription.
For $\rho^2>0$, the NC phase can worsen the UV property of loop integrals
instead of always improving it in high dimensions.
We explain how this surprising phenomenon
comes about from the indefinite metric.
This lends a strong support to the point of view that the naive approach
is not well-founded when time does not commute with space.
For $\rho^2<0$, the NC phase improves the UV property and softens the
quadratic UV divergence in ordinary theory to a bounded but indefinite
UV oscillation. We employ a cut-off method to quantify the new UV
non-regular terms. For $\rho^2>0$, these terms are
generally complex and thus also harm unitarity in addition to those found
previously. As the new terms for both cases are not available
in the Lagrangian and in addition can be
non-Hermitian when time does not commute with space, our result casts
doubts on previous demonstrations of one loop renormalizability based
exclusively upon analysis of planar diagrams, especially in theories with
quadratic divergences.

\begin{flushleft}


\end{flushleft}

\newpage
\section{Introduction}

Quantum field theory on noncommutative (NC) spacetime may be formulated
on ordinary commutative spacetime through the Weyl-Moyal
correspondence, with the point-wise product of field operators in the action
replaced by the Moyal star product,
\begin{equation}
\begin{array}{rcl}
(f_1\star f_2)(x)&=&\displaystyle
\left[\exp\left(\frac{i}{2}
\theta^{\mu\nu}\partial^x_{\mu}\partial^y_{\nu}\right)
f_1(x)f_2(y)\right]_{y=x},
\end{array}
\end{equation}
where $x,~y$ are ordinary commutative coordinates and $\theta_{\mu\nu}$ is
the NC parameter matrix characterizing the NC spacetime,
$[\hat{x}_{\mu},\hat{x}_{\nu}]=i\theta_{\mu\nu}$.

The star product induces a momentum-dependent phase in vertices of Feynman
diagrams in momentum space, which could potentially improve the convergence
property of loop diagrams. However, there always remain a group of diagrams,
i.e., planar ones, that are only modified by a global phase that does not
touch a loop momentum. For these diagrams, the loop integrals are exactly
the same as their ordinary counterparts and thus the ultraviolet
divergences persist $\cite{filk,chaichian}$. More interesting are the
diagrams that do couple external momenta to loop momenta via the NC phases.
These nonplanar (NP) diagrams, as usually called,
exhibit some exotic features that are unfamiliar in ordinary field
theories. For instance, they break unitarity of the S matrix when time is
involved in noncommutativity $\cite{unitarity}$. Nevertheless, there seems
to be no doubt that they are ultraviolet finite at one loop level, barring the
ultraviolet/infrared (UV/IR) mixing problem that happens for exceptional
kinematic configurations $\cite{mixing}$. Then, at least at one loop level,
the fate of renormalizability of an NC field theory rests on the consistent
removal of UV divergences in planar diagrams. Based on this understanding,
it has been extensively demonstrated in the literature that one loop
renormalizability is guaranteed for the NC version of popular commutative
theories; see Refs. $\cite{u1,un}$, for instance, for the first few examples
about $U(1)$ and $U(N)$ gauge theories.

In this work, we examine this statement about finiteness of one loop NP
diagrams. The issue turns out to be much more delicate than expected.
An subtlety appears already in NC theories on Euclidean spacetime:
whether an NP integral converges or not can depend on how we evaluate it.
As this essentially mathematical subtlety does not seem to be solvable by
mathematics, we suggest in section 2 a physical motivation to fix it in
the context of quantum field theory. The situation on Minkowski spacetime
is more involved due to its indefinite metric. To prepare for later
discussions, we review in section 3 some prescriptions that may be used to
evaluate loop integrals in ordinary theory. They are equivalent in the
sense that they always yield the same answer. Then, we present our main
results in section 4, based on the defining prescription used in Feynman
propagator. These results are quite different concerning convergence property from
the conventional ones. The other prescriptions are examined in section 5.
We show that they are generally not equivalent to the defining prescription
any longer, even if they sometimes seem to make sense mathematically.
For an illustration of physical implications of the results in section 4,
we compute in section 6 the one loop NP contributions to the self-energy
and vertex in real $\varphi^4$ theory, and discuss briefly some new points
about unitarity and renormalization. Finally, we summarize our results in
the last section.

\section{Euclidean spacetime}

We consider in this section the simpler case of Euclidean spacetime where
the integrand of a loop integral is real and positive definite before the
NC phase is introduced. The integral of an NP diagram at one loop in $n$
dimensions is essentially of the type,
\begin{equation}
\begin{array}{rcl}
A_{\delta}^n(\rho^2,m^2)&=&\displaystyle
\int\frac{d^nk}{(2\pi)^n}\frac{\cos(k\cdot\rho)}{(k^2+m^2)^{\delta}},
{\rm ~integer~}n\ge 2,{\rm ~integer~}\delta\ge 1,
\end{array}
\label{eq_def1}
\end{equation}
where $m^2>0$, $k^2=k_{\mu}k_{\mu}\ge 0$, $\rho^2=\rho_{\mu}\rho_{\mu}\ge 0$
and $k\cdot\rho=k_{\mu}\rho_{\mu}$.
Here $\rho_{\mu}=\theta_{\mu\nu}p_{\nu}$ with $p$ a linear combination of the
external momenta. For a non-tadpole diagram, one first uses
Feynman parameters to combine propagators and makes a shift in the loop
momentum. In that case, $m^2$ is actually a linear combination of the
internal masses squared and the quadrature of the external momenta, and
those Feynman parameters have to be integrated over upon finishing the above
$k$ integral. The relevant points for us are that $m^2$ is still positive
definite and that introducing Feynman parameter integrals does not
modify UV properties of the original loop integral.
For a convergent integral it is allowed to shift the loop momentum. For a
nonconvergent one, we assume that it has been regularized.
For integrals that involve the phase $e^{ik\cdot\rho}$, we assume
that it is legitimate to ignore the $\sin(k\cdot\rho)$ term odd in $k$. In
the strict sense this is valid only for convergent integrals. For
non-convergent ones and their regularized versions, we will take their
principal values in the sense that,
$\displaystyle\lim_{\Lambda\to\infty}\int_{-\Lambda}^{+\Lambda}dx~f(x)=0$ for an odd
function $f(x)$ even if
$\displaystyle\int_{-\infty}^{+\infty}dx~f(x)$ does not exist.
Finally, the rotational symmetry in $\rho$ is employed in the above notation.

We only need to evaluate $A_1^n$ since $A_{\delta}^n$ is related to
$A_{\delta-1}^n$ by a derivative in $m^2$. This is obvious for convergent
integrals. For non-convergent ones, it is also true as long as the
regularization method used to define them is independent of $m^2$. This
will be the case in our cut-off method. We will not discuss the UV/IR
mixing problem, thus we assume $\rho^2\ne 0$ in our discussion. Then,
with rescaling $\sqrt{\rho^2}k_{\mu}=x_{\mu}$, we have,
\begin{equation}
\begin{array}{rcl}
A_1^n(\rho^2,m^2)&=&\displaystyle
(2\pi)^{-n}(\rho^2)^{1-\frac{n}{2}}E^n\left(\sqrt{m^2\rho^2}\right),\\
E^n(a)&=&\displaystyle
\int d^nx\frac{\cos(x\cdot\hat{\rho})}{x^2+a^2},
~a>0,
\end{array}
\end{equation}
with $\hat{\rho}$ being the unit vector of $\rho$. To illustrate the
subtleties hidden in the original integral $A_1^n$, we employ several
methods to work out $E^n(a)$. Then, we discuss how to manipulate these
integrals in the context of field theory.

\subsection{Mathematical subtlety in $E^n(a)$}

\subsubsection{Method 1}
Without loss of generality, we choose the axis $x_1$ to be in the
direction $\hat{\rho}$. We finish first the $x_1$ integral to obtain,
\begin{equation}
\begin{array}{rcl}
E^n(a)&=&\displaystyle
\pi\int\prod_{j=2}^n(dx_j)
\frac{\exp\left[-\sqrt{\sum_{j=2}^nx_j^2+a^2}\right]}
{\sqrt{\sum_{j=2}^nx_j^2+a^2}},
\end{array}
\end{equation}
which is obviously convergent. Finishing the angular and radial
integration, we have,
\begin{equation}
\begin{array}{rcl}
E^n(a)
&=&\displaystyle\frac{\pi}{2}\Omega_{n-1}a^{n-2}
\int_0^{\infty}dx~x^{\frac{n-3}{2}}
\frac{e^{-a\sqrt{x+1}}}{\sqrt{x+1}}\\
&=&\displaystyle
(2\pi)^{\frac{n}{2}}a^{\frac{n}{2}-1}K_{1-\frac{n}{2}}(a),
\end{array}
\end{equation}
where $\Omega_{\ell}$ is the solid angle in $\ell$ dimensions and
$K_{\nu}(z)$ is the modified Bessel function of order $\nu$ $\cite{mathbook}$,
so that
\begin{equation}
\begin{array}{rcl}
A_1^n(\rho^2,m^2)&=&\displaystyle
(2\pi)^{-\frac{n}{2}}(m^2)^{\frac{n}{2}-1}
\left(\sqrt{m^2\rho^2}\right)^{1-\frac{n}{2}}
K_{1-\frac{n}{2}}\left(\sqrt{m^2\rho^2}\right)
\end{array}
\end{equation}
According to this method, the above result holds in any dimensions.
This coincides with the conventional one in the literature.

\subsubsection{Method 2}
In this method, we reverse the integration order in method 1 to finish first
the subintegral over components perpendicular to the direction $\hat{\rho}$.
It is obvious that for $n\ge 3$ the subintegral diverges so that the
integral itself is not well defined, and that for $n=2$ the result in method 1
is reproduced. These two methods make explicit reference to the direction
defined by $\hat{\rho}$.

\subsubsection{Method 3}

Noting the rotational symmetry of the integral $E^n(a)$, we finish first
the angular integration in $n$ dimensions. Although any method of
angular integration must lead to the same answer, a judicious choice
of angles facilitates this. We choose $x_1=x\cos\theta_{n-1}$ with
$x=\sqrt{x_{\mu}x_{\mu}}$ and $\theta_{n-1}\in[0,\pi]$ so that,
\begin{equation}
\begin{array}{rcl}
E^n(a)&=&\displaystyle
\Omega_{n-1}\int_0^{\infty}x^{n-1}dx\int_0^{\pi}
\sin^{n-2}\theta_{n-1}d\theta_{n-1}
\frac{\cos(x\cos\theta_{n-1})}{x^2+a^2}\\
&=&\displaystyle
(2\pi)^{\frac{n}{2}}
\int_0^{\infty}dx\frac{x^{\frac{n}{2}}}{x^2+a^2}J_{\frac{n}{2}-1}(x),
\end{array}
\end{equation}
where $J_{\nu}(z)$ is the Bessel function of the first kind of order
$\nu$. The leading term of $J_{\nu}(x)$ in the large $x>0$ limit is,
\begin{equation}
\displaystyle
J_{\nu}(x)\to\sqrt{\frac{2}{\pi x}}(\sim\sin x{\rm ~or~}\sim\cos x)
\label{largex}
\end{equation}
Thus the above integral converges only for $\frac{n}{2}-\frac{1}{2}-2<0$,
i.e., for $n<5$, in which case we obtain,
\begin{equation}
\displaystyle
E^n(a)=(2\pi)^{\frac{n}{2}}a^{\frac{n}{2}-1}K_{\frac{n}{2}-1}(a)
\end{equation}
the same result as in method 1 upon using $K_{-\nu}(z)=K_{\nu}(z)$. For
$n\ge 5$ however, the integral is not well-defined in this third
method.

\subsubsection{Method 4}

The last method is the one using Schwinger parametrization that is popular
in the literature. In Euclidean space, the denominator of the integrand in
$E_1^n$ is positive definite, hence it is justified to exponentiate it as
follows,
\begin{equation}
\begin{array}{rcl}
E^n(a)&=&\displaystyle
\int d^nx~e^{ix\cdot\hat{\rho}}
\int_0^{\infty}d\alpha~e^{-\alpha(x^2+a^2)},
\end{array}
\end{equation}
where the cosine factor is replaced by the phase in the sense of principal 
value mentioned earlier if convergence is not guaranteed. 

To make the exponentiation useful for evaluating the integral, we have to
interchange the integration order of $x$ and $\alpha$, which however is
justified only for (absolutely) convergent, or regularized if not, integrals. 
Suppose we take this for granted to proceed further. Then, complete the 
quadrature in $k$, make a complex shift $x\to x+\frac{i}{2\alpha}\hat{\rho}$ that is
permitted because the integrand is analytic in each complex
plane of $x$. Finishing the $x$ integral yields, 
\begin{equation}
E^n(a)=\displaystyle
\pi^{\frac{n}{2}}\int_0^{\infty}d\alpha~\alpha^{-\frac{n}{2}}
e^{-\alpha a^2-\frac{1}{4\alpha}}
=(2\pi)^{\frac{n}{2}}a^{\frac{n}{2}-1}K_{1-\frac{n}{2}}(a),
\end{equation}
which holds seemingly for any dimensions as in method 1. 

\subsection{Discussions}

The above calculations tell us that NP integrals in two dimensional
Euclidean spacetime are indeed unambiguously finite, barring the UV/IR
mixing problem. However, in three and higher dimensions different
computational methods yield different results for the same integral.
In this respect
we do not count method 4 as a justified one, because the very use of
Schwinger parametrization presumes that the integral is well-defined and
thus it cannot say anything when the integral is not well-defined. On
the other hand, the other three methods are completely justified but still
yield different results. This looks quite puzzling.
This is a new feature of NP diagrams in NC field theory that does not
appear in ordinary field theory and that seems to have not been
realized previously.

The answer to the dilemma is simple: the integral $A_1^n$ as originally
defined in eq. $(\ref{eq_def1})$ is not mathematically well-defined in three
and higher dimensions. To define it properly, we should also specify
its integration path or order.
Method 1 amounts to preferred integration over the
component in the $\hat{\rho}$ direction. As the phase is active only in
this direction, this manipulation earns the best convergence property for
the remaining component integrals. Method 2 is just the opposite.
It starts with the components that are independent of the phase.
The subintegral thus corresponds to an ordinary integral
in $(n-1)$ dimensional Euclidean spacetime and divergence starts to appear
at this stage for $n-1\ge 2$. This method gives the worst convergence property.
Method 3 starts with integrating spherically
over directions in $n$ dimensions for a fixed radius and integrating over
the radius later. This produces a mild convergence property.

The result of method 1 can also be understood from another angle, which
explains clearly why the similar subtlety does not appear in ordinary
theory. Take $E^5(a)$ as an example which converges according to method
1 but does not according to methods 2 and 3. We employ the NC phase $\cos x_1$
to do integration by parts repeatedly, reducing the UV divergence
degree by one each time, until we arrive at an integral that converges by
power counting. Since the denominator is positive definite in Euclidean
spacetime, the surface terms vanish. For $E^5(a)$, we need to do so three
times,
\begin{equation}
\begin{array}{rcl}
E^5(a)&=&\displaystyle
24\int\prod_{j\ne 1}(dx_j)\int dx_1~\sin x_1
\left[\frac{x_1}{(\sum_{j=1}^5x_j^2+a^2)^3}
-\frac{2x_1^3}{(\sum_{j=1}^5x_j^2+a^2)^4}\right]
\end{array}
\end{equation}
Since the above is now convergent, we can interchange the integration
order to finish $x_{j\ne 1}$ first and $x_1$ afterwards, yielding the same
result as obtained in method 1.

In the context of field theory, we integrate over momentum fluctuations
of virtual particles in loops. When the above subtlety
appears, mathematically speaking, there is no a priori reason as to
which way of integration we should take. But physically, method 3 is
much better motivated than methods 1 and 2 for the following reason.
A single Feynman diagram usually involves
several NP terms that share the same denominator (i.e., $m^2$) from
propagators but differ in the NC phases (i.e., $\rho_{\mu}$) from the
decomposition of the product of vertices.
In method 3, we treat all terms in the same diagram (actually also all
diagrams) on the
same footing with no preference to any of the $\rho_{\mu}$'s. In methods
1 and 2 however, we accumulate contributions of fluctuations in a manner that
varies according to their preferred directions specified by
$\rho_{\mu}$'s, which seems rather unnatural.

Let us consider the critical case of $n=5$ according to method 3.
To regularize the integral,
we impose a spherical cut-off on the original integral,
$k_{\mu}k_{\mu}\le\Lambda^2$, which amounts to a cut-off
$x_{\mu}x_{\mu}\le\kappa^2$ on the integral $E^5(a)$ with
$\kappa=\Lambda\sqrt{\rho^2}$.
We obtain
\begin{equation}
\begin{array}{rcl}
E^5(a)&=&\displaystyle
8\pi^2\int_0^{\kappa}dx\frac{x^2}{x^2+a^2}
\left(\frac{\sin x}{x}-\cos x\right)\\
&=&\displaystyle
-8\pi^2\sin\kappa+(2\pi)^{\frac{5}{2}}a^{\frac{3}{2}}K_{\frac{3}{2}}(a)
+\cdots,
\end{array}
\end{equation}
where the dots stand for terms vanishing in the limit $\Lambda\to\infty$.
Thus,
\begin{equation}
\begin{array}{rcl}
A_1^5(\rho^2,m^2)&=&\displaystyle
-(4\pi^3)^{-1}(\rho^2)^{-\frac{3}{2}}\sin\left(\Lambda\sqrt{\rho^2}\right)\\
&+&\displaystyle
(2\pi)^{-\frac{5}{2}}(m^2)^{\frac{3}{2}}
\left(\sqrt{m^2\rho^2}\right)^{-\frac{3}{2}}
K_{\frac{3}{2}}\left(\sqrt{m^2\rho^2}\right),
\end{array}
\end{equation}
which has a new term compared to the conventional result. It oscillates but
is bounded in the UV. In even higher dimensions, the new terms are no longer
bounded though oscillating in the UV. For instance, we quote the result for
$n=6$ without giving details,
\begin{equation}
\begin{array}{rcl}
A_1^6(\rho^2,m^2)&=&\displaystyle
-2(2\pi)^{-\frac{7}{2}}\sqrt{\Lambda}(\rho^2)^{-\frac{7}{4}}
\cos\left(\Lambda\sqrt{\rho^2}-\frac{3}{4}\pi\right)\\
&+&\displaystyle
(2\pi)^{-3}(m^2)^2\left(\sqrt{m^2\rho^2}\right)^{-2}
K_2\left(\sqrt{m^2\rho^2}\right)
\end{array}
\end{equation}
From the above derivation, we observe that $A_{\delta}^n$ converges
for $n< 4\delta+1$, oscillates at $n= 4\delta+1$, and diverges oscillatorily
for $n> 4\delta+1$. Although convergence of NP loop integrals is improved
over their commutative counterparts, new terms will appear in high enough
dimensions that are non-regular in the UV and cannot be removed by sending
the cutoff $\Lambda$ to infinity.

In any cut-off method, there is always a question concerning when to impose
the cut-off as the method generally breaks rotational or Lorentz invariance.
In the above calculation, the cut-off is imposed after the original integral
has been cast into the form of eq. $(\ref{eq_def1})$. This guarantees
rotational invariance with respect to external momenta contained in $m^2$ and
$\rho^2$. The cut-off method is still not unique, of course. Using the
spherical cut-off as suggested by the physical motivation, any manipulation,
including method 2, will yield the same answer, though method 3 is the easiest
in practice. Mathematically speaking and without affecting rotational
invariance in external momenta, however, we are not forced to employ
the spherical one. For instance, we may impose two independent cut-offs, $\Lambda_{\perp,\parallel}$, on the loop momentum components perpendicular or
parallel to $\hat{\rho}$. The above mathematical subtlety
then appears in the form that the convergence property depends on the
order of taking the limits of $\Lambda_{\perp}\to\infty$ and
$\Lambda_{\parallel}\to\infty$.

\section{Minkowski spacetime: review of ordinary theory}
Minkowski spacetime has an indefinite metric that produces nontrivial
analytic properties in Green's functions. The same metric will cause
special difficulties in NC theory.
Since the Feynman propagator will necessarily be defined on the complex
plane, an NC phase on the real axis ceases to be a pure phase any longer.
This will modify the analytic properties of Green's functions significantly.
Without taking care of
this, we could make manipulations, especially analytic continuation,
that cannot be justified. This indeed happens, as we will show in the
next sections, when we take for granted that some prescriptions
equivalent in ordinary theory are equally justified in NC theory.
To better see the matter, we review in this section these manipulations
for ordinary loop integrals.

\subsection{An apparent problem}

The one loop integrals in ordinary theory in $n$ dimensional Minkowski
spacetime can be cast into the form,
\begin{equation}
\begin{array}{rcl}
iB_{\delta}^n(m^2-i\epsilon)&=&\displaystyle
\int\frac{d^nk}{(2\pi)^n}\frac{1}{(k^2-m^2+i\epsilon)^{\delta}},
{\rm ~integer~}n\ge 2,{\rm ~integer~}\delta\ge 1,~\epsilon=0^+,
\end{array}
\label{eq_def0}
\end{equation}
with the help of Feynman parameters and free shift of the loop momentum.
Now $m^2$, as a linear form of the internal masses squared and the
quadrature of the external momenta, can be positive (below threshold),
zero (on threshold) and negative (above threshold).
For $n\ge 2\delta$, $B_{\delta}^n$ is divergent in the UV region of $k$ where
$|k^2|$ is also large, i.e., with $k$ being deep space-like or time-like.
This divergence is familiar to us. It is essentially
Euclidean in the sense that it can be figured out by power counting as
in Euclidean space without considering
complications from the indefinite metric.
It is handled by regularization. However, there seems to be another
divergence in the above integral, from the UV region where $|k_0|$ and
$|\vec{k}|$ are equally large but $|k^2|$ is small, i.e., the deep
light-like region of $k$. It is possible only for an indefinite metric
and can be seen more clearly by changing the integration variables
to $x_{\pm}=k_0\pm|\vec{k}|$ upon finishing the angular integration,
\begin{equation}
\begin{array}{rcl}
iB_{\delta}^n(m^2-i\epsilon)
&=&\displaystyle\frac{\Omega_{n-1}}{(2\pi)^n}
2\int_{-\infty}^{\infty}dx_-\int_0^{\infty}dx_+
\frac{(x_++x_-)^{n-2}}{(-4x_+x_--m^2+i\epsilon)^{\delta}},
\end{array}
\end{equation}
where the right-hand side is not well-defined for any $\delta$ in the 
region where one of $|x_{\pm}|$ is large but $|x_+x_-|$ is small.
This apparent divergence is often overlooked justifiably by practitioners
of loop calculations because, as detailed below, it does not actually
appear in ordinary theory.
However, it is questionable that it can also be ignored beyond ordinary
field theory, and particularly in NC theory where time does not commute
with space.

\subsection{The defining prescription}

It is instructive to see how the apparent problem is avoided in
ordinary theory. For this, we trace back to the definition of
Feynman propagator that is the only ingredient appearing in the above
integral.
\begin{equation}
\begin{array}{rcl}
i\Delta_F(x-y)&=&\displaystyle
\langle 0|T\left(\varphi(x)\varphi(y)\right)|0\rangle\\
&=&\displaystyle
\theta(x_0-y_0)\langle 0|\varphi(x)\varphi(y)|0\rangle
+\theta(y_0-x_0)\langle 0|\varphi(y)\varphi(x)|0\rangle\\
&=&\displaystyle
\theta(x_0-y_0)i\Delta^{(+)}(x-y)
+\theta(y_0-x_0)i\Delta^{(+)}(y-x),
\end{array}
\end{equation}
where $\varphi$ is a neutral, free scalar field with mass $m$ and
$|0\rangle$ is the free vacuum state. It has been expressed in terms of
the $i\Delta^{(+)}$ function which itself is defined as the commutator
between the positive- and negative-frequency parts of the field. In four
dimensions, it is
\begin{equation}
\begin{array}{rcl}
i\Delta^{(+)}(x)&=&\displaystyle
\frac{1}{(2\pi)^3}\int\frac{d^3\vec{k}}{2k_0}e^{-ik\cdot x},
~k_0=\sqrt{\vec{k}^2+m^2}\ge m
\end{array}
\end{equation}

Using an integral expression for the step function, the above can be
cast into a four-dimensional integral,
\begin{equation}
\begin{array}{rcl}
i\Delta_F(x)&=&\displaystyle
\int_{C_F}\frac{d^4k}{(2\pi)^4}\frac{i}{k^2-m^2}e^{-ik\cdot x}
\end{array}
\end{equation}
where $C_F$ is a contour in the complex $k_0$ plane. It starts from
$-\infty$ on the real axis, moving below the pole at $k_0=-\sqrt{\vec{k}^2+m^2}$ 
and above the pole at $k_0=+\sqrt{\vec{k}^2+m^2}$, to reach $+\infty$ on the real 
axis, then moves along the semi-circle of infinite radius in the upper
or lower half plane according to $x_0<0$ or $x_0>0$ respectively, to
return back to $-\infty$ on the real axis. The above is usually written in
a concise form using Feynman's $+i\epsilon$ prescription,
\begin{equation}
\begin{array}{rcl}
i\Delta_F(x)&=&\displaystyle
\int\frac{d^4k}{(2\pi)^4}\frac{i}{k^2-m^2+i\epsilon}e^{-ik\cdot x}
\end{array}
\end{equation}
where $\epsilon$ is an infinitesimal real positive number and is the 
source of the $+i\epsilon$ appearing in the loop integral $B_{\delta}^n$.
Although the contour notation is suppressed in the above, it should be
understood 
as being enforced. The two poles have been displaced properly, so that
the contour now extends over the whole real axis of $k_0$ from $-\infty$ to
$+\infty$ 
and returns to $-\infty$ via the semi-circle 
described above. Namely, it is understood that the $k_0$ integral is 
always meant to be finished first by contour integration.
This is necessary to keep the above expression consistent with the 
definition of $\Delta_F(x)$: positive-frequencies propagate only for 
positive time $x_0>0$ while negative-frequencies propagate only for
negative time $x_0<0$. The latter cannot be guaranteed without
enforcing the contour integration.
Furthermore, if the contour integration were not taken into account properly,
$\Delta_F$ and thus $B_{\delta}^n$ would be afflicted with the apparent
problem discussed in subsection 3.1.
This point is often not emphasized in the commutative theory since the
prescription is equivalent to the one of Wick rotation that is easier to
implement.

With the above understanding of Feynman propagator in its manifestly
covariant form, our loop integral becomes using residue theorem,
\begin{equation}
\begin{array}{rcl}
iB_{\delta}^n(m^2-i\epsilon)&=&\displaystyle
\int\frac{d^{n-1}\vec{k}}{(2\pi)^{n-1}}
\int_{C_F}\frac{dk_0}{2\pi}
\frac{1}{[k_0^2-(\vec{k}^2+m^2-i\epsilon)]^{\delta}}\\
&=&\displaystyle i(-1)^{\delta}
\frac{\Gamma(2\delta-1)}{2^{2\delta-1}\left[\Gamma(\delta)\right]^2}
\int\frac{d^{n-1}\vec{k}}{(2\pi)^{n-1}}
(\vec{k}^2+m^2-i\epsilon)^{\frac{1}{2}-\delta},
\end{array}
\end{equation}
where indeed there is no apparent divergence problem that we worried
about earlier.
Here the contour $C_F$ can be closed either in the upper
or in the lower half plane.
Finishing the angular integration in $(n-1)$ dimensions,
then $|\vec{k}|$, and using product theorem for the $\Gamma$ function,
we obtain,
\begin{equation}
\begin{array}{rcl}
B_{\delta}^n(m^2-i\epsilon)&=&\displaystyle
(-1)^{\delta}
\frac{\Gamma(2\delta-1)}{2^{2\delta}\left[\Gamma(\delta)\right]^2}
\frac{\Omega_{n-1}}{(2\pi)^{n-1}}
\int_0^{\infty}dx
\frac{x^{\frac{n-3}{2}}}{(x+m^2-i\epsilon)^{\delta-\frac{1}{2}}}\\
&=&\displaystyle
(-1)^{\delta}
\frac{\Gamma(2\delta-1)}{2^{2\delta}\left[\Gamma(\delta)\right]^2}
\frac{\Omega_{n-1}}{(2\pi)^{n-1}}
(m^2-i\epsilon)^{\frac{n}{2}-\delta}
\frac{\Gamma\left(\delta-\frac{n}{2}\right)
\Gamma\left(\frac{n-1}{2}\right)}
{\Gamma\left(\delta-\frac{1}{2}\right)}\\
&=&\displaystyle
(-1)^{\delta}\frac{\Gamma\left(\delta-\frac{n}{2}\right)}
{\Gamma(\delta)}(4\pi)^{-\frac{n}{2}}
(m^2-i\epsilon)^{\frac{n}{2}-\delta}, 
\end{array}
\end{equation}
where we have assumed $n<2\delta$ to make the integral convergent.
Divergent integrals can be handled as usual.

\subsection{Wick rotation}

As long as we know the apparent problem in subsection $3.1$ does not
actually appear, we can employ other prescriptions that can lead to
the same result for loop integrals.
The only thing that we should take care of is that we 
must not change the analytic properties of the original integral.
Amongst those prescriptions Wick rotation is most appealing since
it can be understood as analytic continuation from Minkowski to
Euclidean spacetime. Namely, the $k_0$ integral can be considered as
being integrated over the imaginary axis in the complex plane of 
$k_0$. Since the 
poles of $k_0$ are located in the second and fourth quadrant, the
rotation has to be counter-clockwise, $k_0\to e^{i\frac{\pi}{2}}k_n$, with
$k_n$ to be integrated over the real axis from $-\infty$ to $\infty$.
The integral can be readily finished since we are now in Euclidean 
loop momentum space,
\begin{equation}
\begin{array}{rcl}
iB_{\delta}^n(m^2-i\epsilon)&=&\displaystyle
e^{i\frac{\pi}{2}\left(1-2\delta\right)}\int\frac{d^nk_E}{(2\pi)^n}
\frac{1}{(k_E^2+m^2-i\epsilon)^{\delta}}\\
&=&\displaystyle
i(-1)^{\delta}\frac{\Omega_n}{(2\pi)^n}
\frac{1}{2}\int_0^{\infty}dx
\frac{x^{\frac{n-2}{2}}}{(x+m^2-i\epsilon)^{\delta}}\\
&=&\displaystyle
i(-1)^{\delta}\frac{\Omega_n}{(2\pi)^n}
\frac{1}{2}(m^2-i\epsilon)^{\frac{n}{2}-\delta}
\frac{\Gamma\left(\delta-\frac{n}{2}\right)
\Gamma\left(\frac{n}{2}\right)}{\Gamma(\delta)},
\end{array}
\end{equation}
identical to the result according to the defining prescription.
This prescription is indeed easier to implement than the first.

\subsection{$\epsilon$ parametrization}
The third prescription is built in the Schwinger parametrization
based on the positive infinitesimal $\epsilon$, here named
$\epsilon$ parametrization for simplicity. Relying on $\epsilon>0$, it is
legitimate to exponentiate the denominator in $B_{\delta}^n$,
\begin{equation}
\begin{array}{rcl}
iB_{\delta}^n(m^2-i\epsilon)&=&\displaystyle
\frac{1}{i^{\delta}\Gamma(\delta)}\int\frac{d^nk}{(2\pi)^n}
\int_0^{\infty}d\alpha~\alpha^{\delta-1}
e^{-\alpha[\epsilon-i(k^2-m^2)]}
\end{array}
\end{equation}
To employ the exponentiation, we have to interchange
the integration order. However, this is permitted only for a
well-defined integral. Here we are not concerned with Euclidean
divergences by power counting, since they can always be handled by
regularization. Instead, we worry about the potential problem 
discussed in subsection $3.1$. Interchanging the integration order
presumes that the integral is indeed free of such a danger, 
and thus has to be justified. This we will do later. 

The $k$ integral can now be finished trivially to yield, 
\begin{equation}
\begin{array}{rcl}
B_{\delta}^n(m^2-i\epsilon)&=&\displaystyle
(4\pi)^{-\frac{n}{2}}i^{-\delta}e^{-i\frac{\pi}{4}n}
\frac{1}{\Gamma(\delta)}
\int_0^{\infty}d\alpha~\alpha^{\delta-\frac{n}{2}-1}
e^{-\alpha(\epsilon+im^2)}
\end{array}
\end{equation}
Note that the Euclidean UV property of the original integral has
been translated into the property of the parameter integral at
$\alpha\to 0$.
The apparent problem that was avoided in the defining prescription by
finishing first the $k_0$ contour integration or in the Wick rotation
by integrating over the imaginary axis of $k_0$,
is now hidden in the behaviour at $\alpha\to\infty$ which is tamed by
$\epsilon>0$. The $\alpha$ integral can be finished straightforwardly to
give the same answer as the previous prescriptions. But this is
not yet what we want, because the above interchange of order has
to be justified. In addition, it is rather disturbing that
convergence at $\alpha\to\infty$ relies on an infinitesimal $\epsilon$.
We recall that $\epsilon$ was introduced only as a short-cut for routing
around the poles and its magnitude should be irrelevant. In fact, 
these two matters are the two faces of the same coin. Namely,
the feasibility of the $\epsilon$ parametrization relies on the
possibility that dependence of convergence on $\epsilon$ can be relaxed.

The integrand of the $\alpha$ integral is analytic in its complex
plane, with a cut along the negative real axis when $n$ is odd.
For $m^2>0$, we can rotate the integration path to the negative
imaginary axis, $\alpha\to e^{-i\frac{\pi}{2}}\beta$ with $\beta\in[0,\infty)$,
because the exponent has a negative real part in the fourth quadrant to
offer an exponential decay in the contribution from the quarter-circle at
infinity:
\begin{equation}
\begin{array}{rcl}
B_{\delta}^n(m^2-i\epsilon)&=&\displaystyle
(4\pi)^{-\frac{n}{2}}(-1)^{\delta}\frac{1}{\Gamma(\delta)}
\int_0^{\infty}d\beta~\beta^{\delta-\frac{n}{2}-1}
e^{-\beta(m^2-i\epsilon)}
\end{array}
\end{equation}
whose outcome is the previous result in the case $m^2>0$. 
For $m^2<0$, we rotate the integration path to the positive 
imaginary axis, $\alpha\to e^{i\frac{\pi}{2}}\beta$. Again, the exponent
provides a sufficient decay for the contribution of the quarter-circle
in the first quadrant. The result is,
\begin{equation}
\begin{array}{rcl}
B_{\delta}^n(m^2-i\epsilon)&=&\displaystyle
(4\pi)^{-\frac{n}{2}}e^{-i\frac{n}{2}\pi}\frac{1}{\Gamma(\delta)}
\int_0^{\infty}d\beta~\beta^{\delta-\frac{n}{2}-1}
e^{-\beta(|m^2|+i\epsilon)}
\end{array}
\end{equation}
which indeed reproduces the previous result for $m^2<0$ by noting
that $(-|m^2|-i\epsilon)^{\frac{n}{2}-\delta}
=e^{i\pi\left(\delta-\frac{n}{2}\right)}|m^2|^{\frac{n}{2}-\delta}$.
The case $m^2=0$ is not a problem either. Remember that we have to
integrate over Feynman parameters hidden in $m^2$. It is dangerous
only when $m^2=0$ occurs at the ends of the parameters, where
it corresponds to a real infrared divergence. A regulator for the
latter, for example, a small mass for the massless particle, also
offers a regulator for $\alpha\to\infty$.

\subsection{Rotation in spatial loop momentum}
Finally, we discuss another prescription that seems to have not
been applied in ordinary field theory calculations. As it leads
to the same result as the above prescriptions for loop integrals
in ordinary theory, it might also offer a possible means in NC
theory. We first finish the $n-1$ dimensional angular integral,
\begin{equation}
\begin{array}{rcl}
iB_{\delta}^n(m^2-i\epsilon)&=&\displaystyle
\frac{\Omega_{n-1}}{(2\pi)^n}\int_{-\infty}^{\infty}dk_0
\int_0^{\infty}dk\frac{k^{n-2}}{(k_0^2-k^2-m^2+i\epsilon)^{\delta}}
\end{array}
\label{eq3.5a}
\end{equation}
with $k=|\vec{k}|$. The integrand of the $k$ integral, considered as a
function on the complex $k$ plane, has two poles of order $\delta$ in
the first and third quadrant. Thus, we may rotate clockwise the
integration path from the real positive axis to the negative
imaginary axis, $k\to e^{-i\frac{\pi}{2}}k$,
\begin{equation}
\begin{array}{rcl}
iB_{\delta}^n(m^2-i\epsilon)&=&\displaystyle
\frac{\Omega_{n-1}}{(2\pi)^n}e^{-i\frac{\pi}{2}(n-1)}
\int_{-\infty}^{\infty}dk_0\int_0^{\infty}dk
\frac{k^{n-2}}{(k_0^2+k^2-m^2+i\epsilon)^{\delta}}\\
\end{array}
\end{equation}
Now write it back to an $n$ dimensional Euclidean integral,
\begin{equation}
\begin{array}{rcl}
iB_{\delta}^n(m^2-i\epsilon)&=&\displaystyle
e^{-i\frac{\pi}{2}(n-1)}\int\frac{d^nk_E}{(2\pi)^n}
\frac{1}{(k_E^2-m^2+i\epsilon)^{\delta}}\\
&=&\displaystyle
e^{-i\frac{\pi}{2}(n-1)}\frac{\Omega_n}{(2\pi)^n}
\frac{1}{2}\int_0^{\infty}dx
\frac{x^{\frac{n-2}{2}}}{(x-m^2+i\epsilon)^{\delta}}
\end{array}
\end{equation}
It can be worked out to reproduce the result given by the other
three prescriptions.

Note the $k$ factor in the numerator in eq. $(\ref{eq3.5a})$. The rotation
looks awkward when $n$ is large enough to deny the justification of ignoring
the contribution from the quarter-circle while Wick rotation in $k_0$
still works.
In such a case, we are not really considering a rotation. Instead, we are
considering whether it is still possible to mimic the result obtained in
the other prescriptions by integrating directly over the imaginary axis of
$|\vec{k}|$. Using the conception of rotation helps us find which half of the
imaginary axis would be appropriate for this.

\subsection{Summary}
It is interesting that the above three manipulations are equivalent
in ordinary theory to the defining one implied in Feynman propagator
as far as the loop integral calculation is concerned.
This feature is due to the simple rational structure of
the integrand which has only the pole singularity and decays as a power
for large loop momentum in all regions except the one close to the
light-like domain. Beyond the ordinary theory, there is no guarantee
that the three manipulations are still justified and should be
equivalent to the defining prescription even if they look mathematically
meaningful when applied by brute force.
But it is clear from the above discussion that, as long as
the manifestly covariant Feynman propagator appears as an ingredient of
the theory, for consistency reasons we should stick to its fundamental
defining prescription when evaluating loop integrals. This is the spirit
that we will follow in the next section. Whether the derived prescriptions
are still justified or equivalent arises as a curious subject that we will
treat later on.

\section{Minkowski spacetime: noncommutative theory}
The loop integral of an NP diagram becomes more involved in Minkowski
spacetime due to the indefinite metric which makes the calculations
in Euclidean spacetime not directly applicable. In this section, we describe
our calculations based on the defining prescription of Feynman propagator
and following the lessons we learned in Euclidean spacetime.
The potential problem discussed in subsection $3.1$ is avoided as in
ordinary theory.

\subsection{Kinematic analysis}

As in Euclidean spacetime, the loop integral of an NP diagram in $n$
dimensional Minkowski spacetime can be reduced to the following form,
\begin{equation}
\begin{array}{rcl}
iB_{\delta}^n(\rho^2,m^2-i\epsilon)&=&\displaystyle
\int\frac{d^nk}{(2\pi)^n}\frac{\cos(k\cdot\rho)}
{(k^2-m^2+i\epsilon)^{\delta}},
{\rm ~integer~}n\ge 2,{\rm ~integer~}\delta\ge 1,~\epsilon=0^+
\end{array}
\label{eq_def2}
\end{equation}
But now $\rho^2$ can take positive (time-like), zero (light-like) and negative
(space-like) values. For instance, for pure space-space noncommutativity,
$\rho^2$ is negative semi-definite; if time is involved, $\rho^2$ can take
either sign.

As we shall see below, the convergence property of the integral
$B_{\delta}^n$ differs according to the sign of $\rho^2$. Thus, to prepare
for the later calculation, we first rewrite the integral in a standard form.
A change of integration variables, a Lorentz transformation $L$ on $k$
in particular, does not change the value of the integral. This leaves
the denominator and integration measure invariant and can be absorbed
into the inverse transformation of $\rho$ in the phase factor.
Note that this may cause a problem for a nonconvergent integral when its
regularized version is not Lorentz invariant. This will be discussed later.
Therefore, by choosing $L$ properly, a $\rho$ with $\rho^2>0$ can be
transformed into a new one with
$\rho_0={\rm sign}(\rho_0)\sqrt{\rho^2}\ne 0,~\vec{\rho}=0$,
and a $\rho$ with $\rho^2<0$ can be transformed into one with
$\rho_0=0,~\vec{\rho}\ne 0,~|\vec{\rho}|=\sqrt{-\rho^2}$.
For $\rho^2=0$, we can make its
components as small as we want but cannot reach the zero limit with a
regular Lorentz transformation, while leaving the value of the integral
unchanged. This is the UV/IR mixing problem when the integral $B_{\delta}^n$
without the phase does not converge. We will not consider this last
case. We emphasize that for the above standardization we need not
assume that $\rho$ is a Lorentz vector. If it is, $\rho^2$ is the same in
all frames. If it is not, it is not a problem either: all components
of $\rho$ must be specified in a given frame before integration, and
then according to this specified $\rho$ we choose our $L$ to make $\rho$
standard. The only difference for us is that a single calculation
holds for all frames in the former case, while we have to do the
calculation frame by frame in the latter case according to the uniquely
specified $\rho$.

As in the case of Euclidean space, $B_{\delta}^n$ and $B_{\delta-1}^n$ are
related by a derivative in $m^2$ when both converge, or when one or
both of them are regularized in a way that does not depend on $m^2$.
We will thus concentrate on $B_1^n$ and discuss briefly the results for
$B^n_{\delta}$.

\subsection{Space-like case: $\rho^2<0$}

The standardized integral is
\begin{equation}
\begin{array}{rcl}
iB_1^n(\rho^2,m^2-i\epsilon)&=&\displaystyle
\int\frac{d^nk}{(2\pi)^n}\frac{\cos(\vec{k}\cdot\vec{\rho})}
{k^2-m^2+i\epsilon},
~|\vec{\rho}|=\sqrt{-\rho^2},
\end{array}
\label{eq4.2.a}
\end{equation}
where only the $(n-1)$ dimensional spatial components of $k$ are
involved in the phase.
Note that it is not possible at this stage to do integration by parts in
the variable $\vec{k}\cdot\hat{\rho}$ as we did in method 1 on Euclidean
spacetime. This is due to the indefinite metric which cannot guarantee
vanishing of the surface terms. This is precisely the same problem we
discussed in subsection $3.1$.
According to the defining prescription of Feynman propagator, the $k_0$
integral should be finished first using residue theorem for the prescribed
contour closed in the upper or lower half complex plane of $k_0$,
\begin{equation}
\begin{array}{rcl}
B_1^n(\rho^2,m^2-i\epsilon)&=&\displaystyle
-\frac{1}{2}\int\frac{d^{n-1}\vec{k}}{(2\pi)^{n-1}}
\frac{\cos(\vec{k}\cdot\vec{\rho})}
{\sqrt{\vec{k}^2+m^2-i\epsilon}}
\end{array}
\label{eq4.2.b}
\end{equation}

We emphasize that the integration order matters here. With the $k_0$
contour integral done first, it is understood that it is finished with
respect to a fixed and finite $|\vec{k}|$ so that one of the two poles stays
always inside the contour. Without enforcing this, the $k_0$ and $\vec{k}$
subintegrals would be interwound since there would be a domain where both
poles stay outside the contour when, roughly speaking, $|\vec{k}|$ is larger
than the radius of the contour.
It is also in this limited sense of the integral that we can consider Wick
rotation (see subsection 5.1) and argue meaningfully that the contribution
from the semi-circle of large radius can be ignored when the integrand
drops fast enough for large $|k_0|$ but finite $|\vec{k}|$.

Now the Euclidean subtlety just mentioned sets in, as it is mathematically
possible to do integration by parts in $\vec{k}\cdot\hat{\rho}$ and ignore the
surface terms, followed by integration over perpendicular components, or
vice versa. The two manipulations yield the best and the worst convergence
respectively. As in the Euclidean case, we have to appeal to the physics
context of field theory where the integral appears, to argue against
manipulations with reference to the direction specified by the $\vec{\rho}$.
This is the only subtlety in calculations using the defining prescription
and it seems unavoidable.

The angular integration yields,
\begin{equation}
\begin{array}{rcl}
B_1^n(\rho^2,m^2-i\epsilon)&=&\displaystyle
-\frac{1}{2}(2\pi)^{\frac{1-n}{2}}\int_0^{\infty}dk~k^{n-2}
\frac{\left(k|\vec{\rho}|\right)^{\frac{3-n}{2}}
J_{\frac{n-3}{2}}\left(k|\vec{\rho}|\right)}
{\sqrt{k^2+m^2-i\epsilon}}
\end{array}
\end{equation}
Using eq. $(\ref{largex})$, we see that $B_1^n$ is not well-defined in the
UV for $n\ge 4$.
Note the difference from the Euclidean case concerning convergence at $n=4$.
There, the angular averaging is over $n$ dimensions, while here it is over
$n-1$ dimensions since $k_0$ has been first finished.
Generally, $B_{\delta}^n$ converges for $n<4\delta$,
oscillates in the UV at $n=4\delta$ and diverges oscillatorily for
$n>4\delta$. As naively expected, the NC phase in the case $\rho^2<0$
indeed improves the convergence property of the loop integral compared to its
commutative counterpart.

To regularize the UV behaviour, we introduce a
spherical cut-off on the spatial loop momentum, $|\vec{k}|\le\Lambda$.
Rescaling $k|\vec{\rho}|=x$ and denoting $\kappa=\Lambda|\vec{\rho}|$, we have,
\begin{equation}
\begin{array}{rcl}
B_1^n(\rho^2,m^2-i\epsilon)&=&\displaystyle
-\frac{1}{2}(2\pi)^{\frac{1-n}{2}}|\vec{\rho}|^{2-n}
\int_0^{\kappa}dx\frac{x^{\frac{n-1}{2}}J_{\frac{n-3}{2}}(x)}
{\sqrt{x^2-a^2-i\epsilon}},
~a^2=m^2\rho^2
\end{array}
\end{equation}
For clarity, we distinguish the two cases of $m^2>0$ (i.e., $a^2<0$)
and $m^2<0$ (i.e., $a^2>0$). The case $m^2=0$ can be covered by either
of them in the limit $m^2\to 0^{\pm}$ since $B_1^n$ has no IR divergence
for $n\ge 2$.
From now on we specialize to four dimensions though there is no
difficulty to work out new, more divergent terms in higher
dimensions. The results for lower dimensions will be discussed in
a separate subsection.

\subsubsection{$m^2>0$}

It is safe to ignore $-i\epsilon$ since the argument of the square
root is positive definite. The non-convergent term can be
separated by integration by parts,
\begin{equation}
\begin{array}{rcl}
-(2\pi)^2|\vec{\rho}|^2B_1^4(\rho^2,m^2-i\epsilon)&=&\displaystyle
\int_0^{\kappa}dx\frac{x\sin x}{\sqrt{x^2+|a^2|}}\\
&=&\displaystyle
-\frac{\kappa\cos\kappa}{\sqrt{\kappa^2+|a^2|}}
+|a^2|\int_0^{\kappa}dx\frac{\cos x}{(x^2+|a^2|)^{\frac{3}{2}}}
\end{array}
\end{equation}
The second term is now well-defined in the limit $\Lambda\to\infty$.
Keeping terms not vanishing in the limit, we have,
\begin{equation}
\begin{array}{rcl}
-(2\pi)^2B_1^4(\rho^2,m^2-i\epsilon)&=&\displaystyle
\frac{1}{\rho^2}\left[\cos\kappa
-\sqrt{|a^2|}K_1\left(\sqrt{|a^2|}\right)\right]
\end{array}
\end{equation}

\subsubsection{$m^2<0$}

Since $a^2>0$, the argument of the square root can become negative 
for small enough $x$ so that we must keep $-i\epsilon$ in that region. 
Note that the integral is well-defined at $x=\sqrt{a^2}$. For the sake
of safety, we split the integral into two parts,
\begin{equation}
\begin{array}{rcl}
-(2\pi)^2|\vec{\rho}|^2B_1^4(\rho^2,m^2-i\epsilon)&=&\displaystyle
i\int_0^{\sqrt{a^2}}dx\frac{x\sin x}{\sqrt{a^2-x^2}}
+\int_{\sqrt{a^2}}^{\kappa}dx\frac{x\sin x}{\sqrt{x^2-a^2}}
\end{array}
\end{equation}
The first integral equals 
$\displaystyle i\frac{\pi}{2}\sqrt{a^2}J_1\left(\sqrt{a^2}\right)$.
The second one is not well-defined in the UV and can be worked out
using the following trick,
\begin{equation}
\begin{array}{rcl}
\displaystyle
\int_{\sqrt{a^2}}^{\kappa}dx\frac{x\sin x}{\sqrt{x^2-a^2}}
&=&\displaystyle
-\int_{\sqrt{a^2}}^{\kappa}\frac{x}{\sqrt{x^2-a^2}}
d\left[\cos x-\cos\sqrt{a^2}\right]\\
&=&\displaystyle
-\frac{\kappa}{\sqrt{\kappa^2-a^2}}
\left[\cos\kappa-\cos\sqrt{a^2}\right]
-a^2\int_{\sqrt{a^2}}^{\kappa}dx\frac{\cos x-\cos\sqrt{a^2}}
{(x^2-a^2)^{\frac{3}{2}}}
\end{array}
\end{equation}
The remaining integral now converges and
equals in the limit $\Lambda\to\infty$,
$$\displaystyle
-\frac{\pi}{2}\sqrt{a^2}N_1\left(\sqrt{a^2}\right)-\cos\sqrt{a^2},$$
where $N_{\nu}(z)$ is the Bessel function
of the second kind of order $\nu$. Collecting terms and keeping only
terms that survive the limit, we have
\begin{equation}
\begin{array}{rcl}
-(2\pi)^2B_1^4(\rho^2,m^2-i\epsilon)&=&\displaystyle
\frac{1}{\rho^2}\left\{\cos\kappa
-i\frac{\pi}{2}\sqrt{a^2}
\left[J_1\left(\sqrt{a^2}\right)+iN_1\left(\sqrt{a^2}\right)\right]
\right\}\\
&=&\displaystyle
\frac{1}{\rho^2}\left[\cos\kappa
-(-i\sqrt{a^2})K_1\left(-i\sqrt{a^2}\right)\right],
\end{array}
\end{equation}
where a relation amongst Bessel functions has been used.

The results for space-like $\rho$ are summarized compactly as
\begin{equation}
\begin{array}{rcl}
-(2\pi)^2B_1^4(\rho^2,m^2-i\epsilon)&=&\displaystyle
\frac{1}{\rho^2}\left[\cos\left(\Lambda\sqrt{-\rho^2}\right)
-z_0K_1(z_0)\right],\\ 
-(2\pi)^2B_2^4(\rho^2,m^2-i\epsilon)&=&\displaystyle
-\frac{1}{2}K_0(z_0),
\end{array}
\label{eq_result1}
\end{equation}
with $z_0=\sqrt{-\rho^2(m^2-i\epsilon)}$. Note that the first term in
$B_1^4$ oscillates in the UV though it is bounded. It is missed in the
conventional calculations. $B_2^4$ and the second term in $B_1^4$
coincide with the conventional results.

\subsection{Time-like case: $\rho^2>0$}

The standardized integral becomes in this case,
\begin{equation}
\begin{array}{rcl}
iB_1^n(\rho^2,m^2-i\epsilon)&=&\displaystyle
\int\frac{d^nk}{(2\pi)^n}\frac{e^{ik_0\rho_0}}{k^2-m^2+i\epsilon},
~\rho_0=\sqrt{\rho^2}>0
\end{array}
\end{equation}
Note that we have used the reflection symmetry to replace
$\cos(k_0\rho_0)$ by the phase in the sense of principal value.
The same symmetry in $k_0$ allows us to take either sign of $\rho_0$.
This affects the choice of contour for the $k_0$ integral accordingly:
for each choice of ${\rm sign}(\rho_0)$, only one of the two defining
contours makes sense; but either choice leads to the same answer of course.
For the above
choice, we close the contour in the upper half complex plane of $k_0$.
Picking up the residue of the pole in the second quadrant and finishing
the trivial angular integration yields,
\begin{equation}
\begin{array}{rcl}
B_1^n(\rho^2,m^2-i\epsilon)&=&\displaystyle
-\frac{1}{2}\frac{\Omega_{n-1}}{(2\pi)^{n-1}}
\int_0^{\infty}dk~k^{n-2}
\frac{e^{-i\rho_0\sqrt{k^2+m^2-i\epsilon}}}
{\sqrt{k^2+m^2-i\epsilon}}
\end{array}
\end{equation}
Taking into account the oscillating phase, the above integral is not
well-defined for $n\ge 3$. Note that there is no similar Euclidean
subtlety as occurring in the case $\rho^2<0$. The above integral is uniquely
specified. One might argue to do integration by parts and ignore
surface terms. This would make any theory free of divergences as it
simply amounts to ignoring any divergences. For the same $|\rho^2|$,
the integral in the time-like case is less convergent than in the
space-like case.
This arises from the asymmetry in the number of separate dimensions in
time and space: while the oscillating behaviour in
the radius $k$ is similar, the angular oscillation occurring in the
space-like is lacking in the time-like case.
This asymmetry disappears in two dimensions.
As above, we introduce a cut-off on the spatial loop momentum
$k\le\Lambda$ to regularize the UV behaviour. Rescaling $k\rho_0=x$ and
denoting $\kappa=\Lambda\sqrt{\rho^2}$, we have,
\begin{equation}
\begin{array}{rcl}
B_1^n(\rho^2,m^2-i\epsilon)&=&\displaystyle
-\frac{1}{2}\frac{\Omega_{n-1}}{(2\pi)^{n-1}}
\left(\sqrt{\rho^2}\right)^{2-n}
\int_0^{\kappa}dx\frac{x^{n-2}e^{-i\sqrt{x^2+a^2-i\epsilon}}}
{\sqrt{x^2+a^2-i\epsilon}},
~a^2=m^2\rho^2
\end{array}
\end{equation}
Below we work out the integral in four dimensions according to the
sign of $m^2$. For a general discussion concerning convergence of
$B_{\delta}^n$, see subsection 4.3.3. For lower dimensional results,
see subsection 4.4.

\subsubsection{$m^2>0$}
In this case $a^2>0$, the integral reduces in four dimensions to
\begin{equation}
\begin{array}{rcl}
-(2\pi)^2\rho^2B_1^4(\rho^2,m^2-i\epsilon)&=&\displaystyle
\int_0^{\kappa}dx\frac{x^2e^{-i\sqrt{x^2+a^2}}}{\sqrt{x^2+a^2}}\\
&=&\displaystyle
i\kappa e^{-i\sqrt{\kappa^2+a^2}}
-i\int_0^{\kappa}dx~e^{-i\sqrt{x^2+a^2}}
\end{array}
\end{equation}
The remaining integral is further reduced in UV divergence as
follows,
\begin{equation}
\begin{array}{rcl}
\displaystyle
\int_0^{\kappa}dx~e^{-i\sqrt{x^2+a^2}}&=&\displaystyle
\int_{\sqrt{a^2}}^{\sqrt{\kappa^2+a^2}}dt
\frac{te^{-it}}{\sqrt{t^2-a^2}}\\
&=&\displaystyle
i\frac{\sqrt{\kappa^2+a^2}}{\kappa}
\left[e^{-i\sqrt{\kappa^2+a^2}}-e^{-i\sqrt{a^2}}\right]
+ia^2\int_{\sqrt{a^2}}^{\sqrt{\kappa^2+a^2}}dt
\frac{e^{-it}-e^{-i\sqrt{a^2}}}{(t^2-a^2)^{\frac{3}{2}}}
\end{array}
\end{equation}
The last term is now well-defined and equals in the limit $\Lambda\to\infty$,
$$\displaystyle
i\left[e^{-i\sqrt{a^2}}-i\sqrt{a^2}K_1\left(i\sqrt{a^2}\right)\right]$$
Thus, we find,
\begin{equation}
\begin{array}{rcl}
-(2\pi)^2B_1^4(\rho^2,m^2-i\epsilon)&=&\displaystyle
\frac{1}{\rho^2}\left[\left(i\kappa+\frac{1}{2}a^2+1\right)e^{-i\kappa}
-i\sqrt{a^2}K_1\left(i\sqrt{a^2}\right)\right],\\
-(2\pi)^2B_2^4(\rho^2,m^2-i\epsilon)&=&\displaystyle
\frac{1}{2}\left[e^{-i\kappa}-K_0\left(i\sqrt{a^2}\right)\right]
\end{array}
\end{equation}

\subsubsection{$m^2<0$}
As we did with the case $\rho^2<0,~m^2<0$, we split the integral into two
parts:
\begin{equation}
\begin{array}{rcl}
-(2\pi)^2\rho^2B_1^4(\rho^2,m^2-i\epsilon)&=&\displaystyle 
i\int_0^{\sqrt{|a^2|}}dx\frac{x^2e^{-\sqrt{|a^2|-x^2}}}{\sqrt{|a^2|-x^2}}
+\int_{\sqrt{|a^2|}}^{\kappa}dx
\frac{x^2e^{-i\sqrt{x^2-|a^2|}}}{\sqrt{x^2-|a^2|}}
\end{array}
\end{equation}
The first term equals
$$
\displaystyle
i\frac{\pi}{2}\sqrt{|a^2|}
\left[I_1\left(\sqrt{|a^2|}\right)-L_1\left(\sqrt{|a^2|}\right)\right],
$$
where $I_{\nu}(z)$ is the other modified Bessel function and $L_{\nu}(z)$ is
the modified Struve function. The second integral is reduced by
integration by parts until we end up with a convergent one in the limit
$\kappa\to\infty$:
\begin{equation}
\begin{array}{rcl}
\displaystyle
\int_{\sqrt{|a^2|}}^{\kappa}dx
\frac{x^2e^{-i\sqrt{x^2-|a^2|}}}{\sqrt{x^2-|a^2|}}&=&\displaystyle
|a^2|\int_0^{\sqrt{\frac{\kappa^2}{|a^2|}-1}}dt~\sqrt{t^2+1}
e^{-it\sqrt{|a^2|}}\\
&=&\displaystyle
\left.\left[i\sqrt{|a^2|}\sqrt{t^2+1}
+\frac{t}{\sqrt{t^2+1}}\right]e^{-it\sqrt{|a^2|}}
\right|_0^{\sqrt{\frac{\kappa^2}{|a^2|}-1}}\\
&-&\displaystyle 
\int_0^{\sqrt{\frac{\kappa^2}{|a^2|}-1}}dt~(t^2+1)^{-\frac{3}{2}}
e^{-it\sqrt{|a^2|}},
\end{array}
\end{equation}
where the last integral is well-defined in the UV and equals
$$
\displaystyle
-\sqrt{|a^2|}K_1\left(\sqrt{|a^2|}\right)
+i\frac{\pi}{2}\sqrt{|a^2|}\left[
L_{-1}\left(\sqrt{|a^2|}\right)-I_1\left(\sqrt{|a^2|}\right)\right]
$$
Collecting terms, keeping only those that survive the limit and
using $\displaystyle L_{-1}(z)-L_1(z)=\frac{2}{\pi}$ to cancel one of the surface
terms, we have,
\begin{equation}
\begin{array}{rcl}
-(2\pi)^2B_1^4(\rho^2,m^2-i\epsilon)&=&\displaystyle
\frac{1}{\rho^2}\left[\left(i\kappa+\frac{1}{2}a^2+1\right)e^{-i\kappa}
-\sqrt{|a^2|}K_1\left(\sqrt{|a^2|}\right)\right],\\
-(2\pi)^2B_2^4(\rho^2,m^2-i\epsilon)&=&\displaystyle
\frac{1}{2}\left[e^{-i\kappa}-K_0\left(\sqrt{|a^2|}\right)\right]
\end{array}
\end{equation}

The results for time-like $\rho$ are summarized as
\begin{equation}
\begin{array}{rcl}
-(2\pi)^2B_1^4(\rho^2,m^2-i\epsilon)&=&\displaystyle
\left(\frac{i\Lambda}{\sqrt{\rho^2}}+\frac{1}{2}m^2+\frac{1}{\rho^2}\right)
e^{-i\Lambda\sqrt{\rho^2}}-\frac{1}{\rho^2}z_0K_1(z_0),\\
-(2\pi)^2B_2^4(\rho^2,m^2-i\epsilon)&=&\displaystyle
\frac{1}{2}\left[e^{-i\Lambda\sqrt{\rho^2}}-K_0(z_0)\right],
\end{array}
\label{eq_result2}
\end{equation}
with $z_0=\sqrt{-\rho^2(m^2-i\epsilon)}$ again. They are indeed very different
from those in the conventional approaches: $B_1^4$ oscillatorily diverges
in the UV while $B_2^4$ oscillates in the UV, though the $K$ function
terms coincide. It is interesting to note that these new terms are complex,
independently of the sign of $m^2$.

\subsubsection{Improving or worsening convergence?}
We saw in the above that the NC phase in the case $\rho^2>0$ softens an
otherwise quadratically divergent integral to an oscillating linearly
divergent one and an otherwise logarithmically divergent integral to an
oscillating but bounded one. However this improvement of convergence is not a
general tendency, but a special feature of the integral $B_1^n$.
To see this, we consider the integral $B_{\delta}^n$ for
general integers $n$ and $\delta>1$ in the case $\rho^2>0$. According to
the defining prescription, we finish first the $k_0$ contour integral by
residue theorem,
\begin{equation}
\begin{array}{rcl}
B^n_{\delta}(\rho^2,m^2-i\epsilon)&=&\displaystyle
\frac{\Omega_{n-1}}{(2\pi)^{n-1}}\frac{1}{\Gamma(\delta)}
\int_0^{\infty}dk~k^{n-2}\\
&\times&\displaystyle
\left[\frac{d^{\delta-1}}{dk_0^{\delta-1}}
\frac{\exp[ik_0\sqrt{\rho^2}]}{(k_0-\sqrt{k^2+m^2-i\epsilon})^{\delta}}
\right]_{k_0=-\sqrt{k^2+m^2-i\epsilon}}
\end{array}
\end{equation}

Now the derivative can act on the denominator and the phase. When it
acts on the phase, it does not lower the divergence degree in $k$ so that
we lose in convergence compared to the ordinary theory. This is a
consequence of the interplay that we referred to at the beginning of
section 3 between the complexation of the phase and the propagator on
the $k_0$ plane, which makes Wick rotation impossible and causes
the result from the contour integration less convergent. It seems
inevitable for the correct manipulation of propagators to avoid the
apparent problem in subsection 3.1. The fate of
convergence thus rests on the term where all derivatives act on the
phase, which is
\begin{equation}
\begin{array}{rcl}
B^n_{\delta}&=&\displaystyle
\frac{\Omega_{n-1}}{(2\pi)^{n-1}}
\frac{[i\sqrt{\rho^2}]^{\delta-1}}{(-2)^{\delta}\Gamma(\delta)}
\int_0^{\infty}dk~k^{n-2}
\frac{\exp\left[-i\sqrt{\rho^2(k^2+m^2-i\epsilon)}\right]}
{\left(\sqrt{k^2+m^2-i\epsilon}\right)^{\delta}}+\cdots,
\end{array}
\end{equation}
where the dots stand for less divergent terms. Using a cut-off $\Lambda$,
the leading divergence is
\begin{equation}
\begin{array}{rcl}
B^n_{\delta}&=&\displaystyle
\frac{\Omega_{n-1}}{(2\pi)^{n-1}}
\frac{[i\sqrt{\rho^2}]^{\delta-1}}{(-2)^{\delta}\Gamma(\delta)}
\frac{i}{\sqrt{\rho^2}}\Lambda^{n-2-\delta}e^{-i\Lambda\sqrt{\rho^2}}
+\cdots
\end{array}
\end{equation}
Thus the integral converges for $n<\delta+2$, oscillates at
$n=\delta+2$ and diverges oscillatorily for $n>\delta+2$.
For comparison, the ordinary integral without a phase converges for
$n<2\delta$, diverges logarithmically at $n=2\delta$ and as a power
$\Lambda^{n-2\delta}$ for $n>2\delta$, with the same UV cut-off.
Thus, only at $\delta=1$, the NC phase improves the convergence property
of the integral by reducing the divergence degree by one.
At $\delta=2$, the divergence in ordinary theory is modulated by an
oscillating phase, except at $n=4$ where a logarithmic divergence softens
to a bounded oscillating factor. For $\delta\ge 3$, convergence can be
worsened in high enough dimensions;
i.e., a convergent integral in ordinary theory
(when $n<2\delta$) is rendered divergent ($n>\delta+2$) or oscillating
($n=\delta+2$) by the phase.
For $\delta>1$, $B_{\delta}^n$ contains $\delta$ terms that successively
differ by one in divergence degree. The terms whose convergence
is improved over the commutative case are singular at $\rho^2=0$, which
is the UV/IR mixing problem, while the terms whose convergence is
worsened or just modulated by a phase are smooth in the limit
$\rho^2\to 0$.

\subsection{Results in two and three dimensions}

For completeness, we include in this subsection the results in lower
dimensions. The two dimensional case is interesting because time and space
are more symmetric than in higher dimensions. Here, we have
$\theta_{\mu\nu}=\theta\epsilon_{\mu\nu}$ with $\epsilon_{\mu\nu}$ being the
Lorentz invariant antisymmetric constant tensor, so that $\rho^2=-\theta^2 p^2$.
The cases of $\rho^2>0$ and $\rho^2<0$ are then related by a sign flip of the
external momentum quadrature and there should be no essential difference
between the two. This indeed coincides with our previous results that
all NP integrals are convergent in two dimensions. Actually, any of the four
prescriptions may be used to evaluate the integral as long as it makes sense
mathematically. In what follows, we establish the above mentioned relation
more directly.

It is sufficient for us to consider $B_1^2$,
\begin{equation}
\begin{array}{rcl}
\displaystyle iB_1^2(\rho^2,m^2-i\epsilon)&=&\displaystyle
\int\frac{d^2k}{(2\pi)^2}
\frac{1}{k^2-m^2+i\epsilon}\times\left\{
\begin{array}{ll}
\cos(k_0\sqrt{\rho^2})&{\rm for~}\rho^2>0\\
&\\
\cos(k_1\sqrt{-\rho^2})&{\rm for~}\rho^2<0
\end{array}\right.
\end{array}
\end{equation}
We relate it to $B_1^2(-\rho^2,-m^2-i\epsilon)$. Complex conjugation and
interchanging the integration variables for the case $\rho^2>0$ yields,
\begin{equation}
\begin{array}{rcl}
\displaystyle -iB_1^{2*}(\rho^2,m^2-i\epsilon)&=&\displaystyle
-\int\frac{d^2k}{(2\pi)^2}
\frac{\cos\left(k_1\sqrt{-(-\rho^2)}\right)}{k^2-(-m^2)+i\epsilon},
\end{array}
\end{equation}
namely,
\begin{equation}
B_1^{2*}(\rho^2,m^2-i\epsilon)=B_1^2(-\rho^2,-m^2-i\epsilon)
\end{equation}
which now holds for any sign of $\rho^2$. Thus, only the relative sign of
$\rho^2$ and $m^2$ is relevant. The explicit result is
\begin{equation}
\displaystyle
-(2\pi)B_1^2(\rho^2,m^2-i\epsilon)
=K_0(z_0)
\end{equation}
Thus $B_1^2$ is real for $\rho^2m^2<0$ and complex for $\rho^2m^2>0$.

The results in three dimensions are
\begin{equation}
\begin{array}{rcl}
\displaystyle
-(2\pi)^{\frac{3}{2}}B_1^3(\rho^2,m^2-i\epsilon)&=&\displaystyle
\frac{\sqrt{m^2-i\epsilon}}{\sqrt{z_0}}K_{\frac{1}{2}}(z_0)
+\left\{
\begin{array}{ll}
0&{\rm for~}\rho^2<0\\
i\sqrt{\frac{\pi}{2\rho^2}}e^{-i\Lambda\sqrt{\rho^2}}
&{\rm for~}\rho^2>0
\end{array}\right.
\end{array}
\end{equation}

\subsection{Further discussions}

Before we started off to evaluate the integral, we first made the integral
standardized. This is all right if it is well-defined, as it amounts to a
change of integration variables. For integrals that are not well-defined
however, this change of variables may clash with the regularization
used to define them. This is indeed the case with the cut-off methods as
we discussed at the end of subsection 2.2 for the Euclidean case.
Since a suitable regularization should maintain Lorentz symmetry (suppose
for simplicity that $\rho$ is a Lorentz vector), all reference to
individual components of $\rho$ must be removed before the
cut-off is imposed. This is the other reason that we first standardized
the integrals, in addition to the one explained in subsection 4.1.
But differently from the Euclidean space, we cannot
regularize the UV behaviour by simply imposing $|k^2|\le\Lambda^2$ due to
the apparent problem exposed in section 3.1. In ordinary theory this is
circumvented by making a Wick rotation beforehand. However, as we explained
previously and will detail in the next section, there is generally no
guarantee for the justification of Wick rotation in NC theory.
Furthermore, in the above approach based on the defining prescription,
we never put a cut-off on $k_0$ as this is demanded by self-consistency of
the Feynman propagator; instead, if necessary, the cut-off is always
imposed on spatial components after finishing the $k_0$ contour integration.
We stress that the new terms in eqs. $(\ref{eq_result1},\ref{eq_result2})$
compared to the conventional results are not regular in the UV and do not
disappear in the limit of $\Lambda\to\infty$.

\section{Ambiguities on noncommutative Minkowski spacetime}
In the preceding section we evaluated the NP integrals according to the
defining prescription built in Feynman propagator. This is the fundamental
prescription that is justified by consistency whenever Feynman propagator
arises as an ingredient of any theory. In that evaluation, only one
subtlety remains for $\rho^2<0$, which is essentially the same one as
in Euclidean spacetime exposed in section 2. As it cannot be excluded
mathematically, it has to be fixed by physical motivations. The results are
thus certain up to this subtlety. Note that the latter appears both for
space-space and time-space noncommutativity.

In this section, we employ the other three prescriptions to calculate the
NP integrals when this seems to make sense mathematically.
Our purpose is to show that equivalent manipulations for ordinary theory
are not necessarily equivalent any more in NC theory.
Ambiguities arise in the sense that these three prescriptions generally
lead to different results concerning convergence than the defining one.
In the course of doing so, it will also become clear how the difference
occurs between the calculation in the preceding section and the
conventional approaches.

\subsection{Wick rotation}
When $\rho^2<0$, $k_0$ is not involved in the NC phase in the integral
$B_1^n$. It thus makes sense mathematically to integrate over the imaginary
axis of $k_0$, which amounts to Wick rotation in $k_0$,
\begin{equation}
\begin{array}{rcl}
B_1^n(\rho^2,m^2-i\epsilon)&=&\displaystyle
-\int\frac{d^nk_E}{(2\pi)^n}
\frac{\cos\left(\vec{k}^{(n-1)}\cdot\vec{\rho}\right)}{k^2_E+m^2-i\epsilon},
\end{array}
\end{equation}
where $\vec{k}^{(n-1)}$ reminds us that the rotated $n$-th component is not
involved in the phase.
The above integral taken in its face is not uniquely specified as different
sequential manipulations
can result in different answers concerning convergence. We will not go into
the details but listing the results.
(1) If we finish $k_n$ first, the result coincides with that in the preceding
section, including the same Euclidean subtlety. This is just as expected,
since we know that
$k_0$ before rotation, corresponding to $k_n$ after rotation, should be finished
first. The coincidence is guaranteed by the analytic properties.
The following three manipulations are similar to the Euclidean case.
(2) If we first do $n$ dimensional angular averaging treating $\vec{\rho}$ as
a vector in $n$ dimensions, the integral converges for $n<5$. This looks not
quite natural. Why should we still do an $n$ dimensional spherical averaging
now that $\vec{\rho}$ has no $n$-th component?
(3) If the component $\vec{k}\cdot\hat{\rho}$ is finished first, the integral
converges for any $n$.
(4) Since $k_E^2\ge 0$, the denominator may be exponentiated by Schwinger
parametrization. Even if $m^2<0$, this is still all right since the $k_E$
domain with the denominator negative is finite and thus the exponentiation
cannot modify the UV property in $k_E$. This is a popular approach in the
literature. However, as explained in subsection 2.1, for the parametrization
to go through, we have to assume that the integral converges. It is thus not
surprising that its outcome converges for any $n$ in this manipulation.

For $\rho^2>0$, there is no way to invoke Wick rotation since the NC phase
blows up on either the positive or negative imaginary axis, depending on
the choice of the phase sign. In Ref. $\cite{unitarity}$, this is
circumvented by simultaneous analytic continuation of the external momenta
$p$ and $\theta_{\mu\nu}$ (and thus $m^2$ and $\rho$) together with $k_0$, so
that a phase on the real axis remains to be a phase on the imaginary axis.
[Note that `simultaneous' continuation is vital to avoid blowing-up
exponentials. There is no step in between, since it does not help to
continue $p$ and $\theta_{\mu\nu}$ before or after continuing $k_0$.
Such a continuation is very unusual indeed. It was also used in Ref.
$\cite{unitarity}$ to argue away a potential UV divergence in NP loop integrals.]
After finishing the integration, the external momenta and $\theta_{\mu\nu}$
are continued back to Minkowski spacetime. However, it is precisely this
continuation forth and back in $p$ and $\theta_{\mu\nu}$
that has to be justified, instead of being assumed.
We know from the above discussion that convergence properties on NC Euclidean
spacetime are very different from those on NC Minkowski spacetime for
$\rho^2>0$ due to their big difference in analytic properties:
in the former, everything else except the phase is positive-definite and thus
the integral can be finished in principle without continuation from the
real axis to the complex plane; in the latter, analytic continuation is
necessary due to the apparent problem associated with the indefinite metric,
which however is made nontrivial by the NC phase.

\subsection{$\epsilon$ parametrization}

This manipulation is more tricky and looks harder to exclude than the other
two. It starts with the justified Schwinger
parametrization based on the infinitesimal $\epsilon>0$:
\begin{equation}
\begin{array}{rcl}
i^{\delta+1}\Gamma(\delta)B_{\delta}^n(\rho^2,m^2-i\epsilon)&=&\displaystyle
\int\frac{d^nk}{(2\pi)^n}e^{ik\cdot\rho}
\int_0^{\infty}d\alpha~\alpha^{\delta-1}
e^{-\alpha\left[\epsilon-i(k^2-m^2)\right]},
\end{array}
\end{equation}
with integer $\delta\ge 1$.
As we will show below, this only postpones the problem instead of solving it.
To proceed further, we have to interchange the integration order of $k$
and $\alpha$ so that we can finish the loop integral. As we stressed repeatedly,
this is legitimate only when the original integral is well-defined.
To see that the problem will come back, let us tolerate this and press on.
Complete the $k$ quadrature and make a shift $k\to k-\rho/(2\alpha)$.
The $k$ integral is finished as in the ordinary case,
\begin{equation}
\begin{array}{rcl}
\displaystyle
i^{\delta}e^{-i\frac{n}{4}\pi}(4\pi)^{\frac{n}{2}}\Gamma(\delta)B_{\delta}^n
&=&\displaystyle\int_0^{\infty}d\alpha~\alpha^{\delta-1-\frac{n}{2}}
\exp\left[-i\alpha(m^2-i\epsilon)-i\frac{\rho^2}{4\alpha}\right]
\end{array}
\label{eq_a}
\end{equation}

Since the $\alpha^{-1}$ term in the exponent is purely imaginary, the above
integral is not well behaved at $\alpha\to 0$ for $n\ge 2(\delta+1)$ and hence
cannot be directly evaluated in terms of Bessel functions. This is the standard
UV divergence though seemingly improved, but it implies already that the NP
integral $B_{\delta}^n$ is not always UV finite.
Furthermore, as in the ordinary case, $\epsilon$ is a routing indicator whose
absolute magnitude should be irrelevant. Namely, to justify the parametrization
posteriorly, dependence of convergence on $\epsilon$ at $\alpha\to\infty$
should be relaxed. We recall that this is the region which causes an apparent
problem due to the indefinite metric in the ordinary case.
Nevertheless, for the most frequently studied diagram in the literature with
two propagators in four dimensions, i.e., $B_2^4$, the above seems to give a
well-defined result even if we set $\epsilon=0$,
\begin{equation}
\begin{array}{rcl}
(4\pi)^2\Gamma(2)B_2^4(\rho^2,m^2)
&=&\displaystyle\int_0^{\infty}d\alpha~\alpha^{-1}
\exp\left[-i\alpha m^2-i\frac{\rho^2}{4\alpha}\right]\\
&=&\displaystyle\left\{
\begin{array}{ll}
2K_0\left({\rm sign}(\rho^2)~i\sqrt{\rho^2m^2}\right)&{\rm for ~}\rho^2m^2>0\\
2K_0\left(\sqrt{|\rho^2m^2|}\right)&{\rm for ~}\rho^2m^2<0
\end{array}\right.,
\end{array}
\end{equation}
which may be summarized concisely as the conventional result,
\begin{equation}
\begin{array}{rcl}
(4\pi)^2\Gamma(2)B_2^4(\rho^2,m^2)
&=&\displaystyle
2K_0\left(\sqrt{-\rho^2(m^2-i\epsilon)}\right)
\end{array}
\end{equation}
However, the above manipulation is rather questionable. We started with a
positive $\epsilon$ whose non-vanishing is crucial for us to apply the Schwinger
parametrization and whose role waits to be relaxed. But finally we found it is
redundant for the convergence of the parameter integral. Although it appears
in the final compact result, this re-introduction is only for the purpose of
bookkeeping. To see the point more definitely, let us apply the same practice
of setting $\epsilon=0$ to the integral $B_3^4$,
\begin{equation}
\begin{array}{rcl}
i(4\pi)^2\Gamma(3)B_3^4(\rho^2,m^2)
&=&\displaystyle\int_0^{\infty}d\alpha~
\exp\left[-i\alpha m^2-i\frac{\rho^2}{4\alpha}\right],
\end{array}
\end{equation}
which obviously diverges at $\alpha\to\infty$. But we know from the last
section using the safest prescription that $B_3^4$ converges for either sign
of $\rho^2$. Thus, the role played by a nonvanishing $\epsilon$ cannot simply
be ignored. The only way left for the feasibility of the method is to
relax the dependence of convergence on $\epsilon$ by rotating $\alpha$ from
the positive real axis to the positive or negative imaginary axis.

We showed in subsection 3.4 that the relaxation is always possible in the
ordinary case, which thus confirms the equivalence of the parametrization to
the defining prescription. In the NC case however, this is not always
possible. For $\rho^2m^2>0$ (i.e., below threshold for $\rho^2>0$ or above
threshold for $\rho^2<0$), the $\alpha$ and $\alpha^{-1}$ terms have an opposite
sign in the real parts for complex $\alpha$. It is thus not possible to
rotate it to either its positive or negative imaginary axis, to avoid
exponential blowing-up. Note that the power factor in eq. $(\ref{eq_a})$ does
not help on this matter. This is strong enough to deny the feasibility of the
approach, though it seems to work for $\rho^2m^2<0$, because, for each
sign of $\rho^2$ both kinematic configurations of $m^2>0$ and $m^2<0$
are permitted and cannot be covered simultaneously.

\subsection{Rotation in $|\vec{k}|$}

For $\rho^2<0$, analytic continuation in $|\vec{k}|$ is not possible:
since we are only allowed to rotate it in the fourth quadrant due to its pole
in the first quadrant, this rotation is allowed only for the
$e^{-i|\vec{k}|\sqrt{-\rho^2}}$ term but not for the
$e^{+i|\vec{k}|\sqrt{-\rho^2}}$ term, both of which appear
after angular integration. For $\rho^2>0$, the rotation is possible
mathematically, up to the explanation in subsection 3.5.
Consider again $B_1^n$, which becomes after angular integration,
\begin{equation}
\begin{array}{rcl}
iB_1^n(\rho^2,m^2-i\epsilon)&=&\displaystyle
\frac{\Omega_{n-1}}{(2\pi)^n}\int_{-\infty}^{\infty}dk_0~\cos(k_0\rho_0)
\int_0^{\infty}\frac{k^{n-2}dk}{k_0^2-k^2-m^2+i\epsilon}
\end{array}
\end{equation}
We rotate $k$ in the fourth quadrant, $k\to e^{-i\frac{\pi}{2}}k$, to move to
the Euclidean spacetime; then we put back the angular integration,
\begin{equation}
\begin{array}{rcl}
B_1^n(\rho^2,m^2-i\epsilon)
&=&\displaystyle
e^{-i\frac{n}{2}\pi}\int\frac{d^nk_E}{(2\pi)^n}
\frac{\cos(k_0\sqrt{\rho^2})}{k_E^2-m^2+i\epsilon}
\end{array}
\end{equation}
Note the opposite sign in the denominator compared to Wick rotation.

Rescaling $k_E\sqrt{\rho^2}=x$ and relabelling $x$, the integral becomes,
\begin{equation}
\begin{array}{rcl}
e^{i\frac{\pi}{2}n}(\rho^2)^{\frac{n}{2}-1}
B_{\delta}^n(\rho^2,m^2-i\epsilon)&=&\displaystyle
\int\frac{d^nx}{(2\pi)^n}\frac{\cos x_1}{x^2-m^2\rho^2+i\epsilon}
\end{array}
\end{equation}
As far as the UV property is concerned, the above integral is the same as that
we treated in Euclidean spacetime. This is obvious for $m^2<0$. When $m^2>0$,
the denominator has a sign flip at a finite value of $x$, which does not modify
the UV property and is taken of by $+i\epsilon$. The integral can be similarly
worked out with same subtleties as appearing there, which we do not repeat.
For comparison, we recall that there is no subtlety for $\rho^2>0$ in the
evaluation using the defining prescription.
All this shows clearly that the prescription of rotation in $|\vec{k}|$
is no more equivalent to the defining one in NC theory.
The weakest point of rotation in
$|\vec{k}|$ is perhaps its lack of physical motivation compared to Wick
rotation, which in ordinary theory relates Minkowski to Euclidean
spacetime.

\section{Application to real $\varphi^4$ theory}

To illustrate physical implications of the above calculations, we consider
briefly the real $\varphi^4$ theory in four dimensions,
\begin{equation}
\begin{array}{rcl}
{\cal L}&=&\displaystyle
\frac{1}{2}\partial_{\mu}\varphi\star\partial^{\mu}\varphi
-\frac{1}{2}m^2\varphi\star\varphi
-\frac{\lambda}{4}\varphi\star\varphi\star\varphi\star\varphi
\end{array}
\end{equation}
The Feynman rule for the vertex is
\begin{equation}
-i2\lambda[c_{12}c_{34}+c_{23}c_{14}+c_{31}c_{24}]
\end{equation}
with $c_{ij}=\cos(p_i\wedge p_j)$, where $p_i$ is the incoming momentum of the
$i$-th particle, and $p\wedge q=\theta_{\mu\nu}p^{\mu}q^{\nu}$.

The NP part of the one-loop self-energy, defined as $i\times$diagram, is
\begin{equation}
\begin{array}{rcl}
\Sigma_{\rm NP}(\tilde{p}^2,m^2-i\epsilon)&=&\displaystyle
-\lambda B_1^4(\tilde{p}^2,m^2-i\epsilon),
\end{array}
\end{equation}
where $p$ is the external momentum.
This diagram is usually not checked probably because it is too simple. But
according to eqs. $(\ref{eq_result1},\ref{eq_result2})$, it is not.
For $\tilde{p}^2>0$, which is possible when $\theta_{0j}\ne 0$, the standard
term proportional to $\displaystyle\frac{1}{\tilde{p}^2}z_0K_1(z_0)$, with
$z_0=\sqrt{-\tilde{p}^2(m^2-i\epsilon)}=i\sqrt{\tilde{p}^2m^2}$, violates
unitarity, though its planar counterpart is empty in unitarity.
The new terms are oscillatorily divergent in the UV. Since they are complex,
they violate unitarity too, and furthermore the violation is even not finite.
For $\tilde{p}^2<0$, the diagram has a new term which is bounded but
oscillates in the UV. Note that the new terms cannot be absorbed by
renormalization of the bilinear terms in ${\cal L}$.

Consider now the one-loop vertex with incoming external momenta
$p_j$. There are three diagrams related by crossing symmetry. It is
sufficient to consider the contribution from the diagram where particles 1
and 2 merge into 3 and 4. After some tedious algebra, the NP part of
the vertex, $\Gamma^{(12)}_{\rm NP}$, defined as $i\times$diagram, is reduced to
\begin{equation}
\begin{array}{rcl}
\displaystyle-\frac{1}{2\lambda^2}
\Gamma^{(12)}_{\rm NP}(p_i,\tilde{p}_j)
&=&\displaystyle\int_0^1dx\left\{
2c_{12}c_{34}B_2^4\left(\rho_{12}^2,A_{12}-i\epsilon\right)\right.\\
&&\displaystyle
+c_{12}\cos((2x-1)p_3\wedge p_4)\left[
B_2^4\left(\rho_3^2,A_{12}-i\epsilon\right)
+B_2^4\left(\rho_4^2,A_{12}-i\epsilon\right)\right]\\
&&\displaystyle
+c_{34}\cos((2x-1)p_1\wedge p_2)\left[
B_2^4\left(\rho_1^2,A_{12}-i\epsilon\right)
+B_2^4\left(\rho_2^2,A_{12}-i\epsilon\right)\right]\\
&&\displaystyle
+\frac{1}{2}\cos((2x-1)(p_1\wedge p_2-p_3\wedge p_4))
B_2^4\left(\rho_{13}^2,A_{12}-i\epsilon\right)\\
&&\displaystyle
+\frac{1}{2}\left.\cos((2x-1)(p_1\wedge p_2+p_3\wedge p_4))
B_2^4\left(\rho_{14}^2,A_{12}-i\epsilon\right)\right\}
\end{array}
\end{equation}
where
$A_{12}=m^2-p_{12}^2x(1-x)$,
$\rho_a^{\mu}=\theta^{\mu}_{~\nu}p_a^{\nu}$ with
$a=1,2,3,4,12,13,14$, and $p_{12}=p_1+p_2$, etc.
When all $\rho_a^2<0$, as is the case with pure space-space noncommutativity,
there is nothing new compared to the conventional results. When some
$\rho_a^2>0$ however, there will be new contributions to the vertex.
We will not attempt here a detailed kinematic analysis to exhaust all
possibilities but giving a few examples. For instance, the separate new term
from $\rho_{12}^2>0$, $\rho_1^2>0$ and $\rho_{13}^2>0$ is
\begin{equation}
\begin{array}{l}
\displaystyle
-(2\pi)^{-2}c_{12}c_{34}e^{-i\Lambda\sqrt{\rho_{12}^2}},\\
\displaystyle
-\frac{1}{2}(2\pi)^{-2}c_{12}\frac{\sin(p_3\wedge p_4)}{p_3\wedge p_4}
e^{-i\Lambda\sqrt{\rho_1^2}},\\
\displaystyle
-\frac{1}{4}(2\pi)^{-2}
\frac{\sin(p_1\wedge p_2-p_3\wedge p_4)}{p_1\wedge p_2-p_3\wedge p_4}
e^{-i\Lambda\sqrt{\rho_{13}^2}}
\end{array}
\end{equation}
Each of these terms oscillates in the UV, is complex thus harming unitarity,
and smooth in the commutative limit.

\section{Conclusions}

We have studied the loop integrals appearing in one loop NP diagrams of
NC field theory and obtained results that are very different from those
in the literature. We found that a mathematical subtlety occurs already on
Euclidean spacetime. We appealed to a physical motivation in the context
of quantum field theory to fix it up. With this manipulation, NP integrals
converge in four and lower dimensions but can diverge or oscillate in the
UV in higher dimensions. We employed a simple cut-off method to illustrate
the new UV non-regular terms that were missed in the conventional approaches.

The situation becomes more involved on Minkowski
spacetime due to an interplay between the indefinite metric and
complexation of the NC phase. For consistency reasons that positive
(negative) frequencies must propagate forward (backward) in time,
the manifestly covariant Feynman propagator has to be defined
as a contour integral on the complex plane. This avoids an apparent
problem due to the indefinite metric in ordinary loop integrals.
We emphasized that it is thus mandatory to evaluate NP integrals of NC theory
in this defining prescription where the zero component of the loop momentum
is first worked out by the contour integration.
It has to be examined as a separate issue
whether other prescriptions that are equivalent in
ordinary theory yield the same result as the defining one.
Indeed, an NC phase defined on the real axis ceases to be a phase on
the complex plane, but can blow up in the upper or lower half plane
when the NC external momentum is time-like, for instance. This directly
blocks up the usage of some of the other prescriptions in the NC case.
The remaining prescriptions, even if they may seem to work mathematically, are
either not physically well-motivated (e.g., rotation in $|\vec{k}|$ in the
case of a time-like external NC momentum), or cannot go through for all
possible kinematic configurations (e.g., $\epsilon$ parametrization), or
can be argued away using the same physical motivation we appealed to on
Euclidean spacetime (e.g., Wick rotation for a space-like NC external
momentum). These discussions show that equivalent prescriptions in ordinary
theory are not necessarily equivalent in NC theory.

An NC phase involving a space-like NC external momentum always improves upon
the convergence property. In four dimensions, such a phase softens an
otherwise quadratically divergent integral to a bounded and UV oscillating
one, while it saves a logarithmic divergence to a finite result (up to
the UV/IR mixing as always). In higher dimensions however, new UV divergent
terms appear that are modulated by an oscillating factor.
When a time-like NC external momentum is involved, the NC phase does not
always improve convergence but can worsen it in high enough dimensions.
This is a surprising result indeed and has its origin in the interplay
discussed above. It lends a strong support to the opinion that Feynman
propagator does not appear as a basic ingredient in perturbative NC theory
when time does not commute with space $\cite{bahns,rim,liao}$ so that the
naive approach built on it is not well-founded.
In four dimensions, such a phase softens a quadratically divergent integral
to a linearly divergent one modulated by an oscillating factor,
and a logarithmically divergent integral to an oscillating one.
We determined those new terms using the cut-off regularization in the
defining prescription. As we demonstrated in the real $\varphi^4$ theory,
they endanger renormalizability already at one loop level, and become
complex in the case of a time-like NC external momentum and thus
harm unitarity as well.

\vspace{0.5cm}
\noindent
{\bf Acknowledgements}

I'm grateful to C. Dehne, P. Heslop, T. Reichenbach and K. Sibold
for many valuable discussions that have helped me refine some arguments,
and M. Chaichian, K. Sibold and A. Tureanu for reading the manuscript
and helpful comments. I'm indebted to Y.-B. Dai for reading the manuscript
and enlightening discussions, and X. Q. Li-Jost for helpful conservations on
section 2. I'd like to thank N. Seiberg for an electronic communication
and T. Mehen for electronic comments on the manuscript.


\end{document}